Rapid change of electronic anisotropy in overdoped $(Y,Ca)Ba_2Cu_3O_{7-\delta}$

K. Nagasao, T. Masui, S. Tajima

Department of Physics, Osaka University, 560-0043, Japan

**Abstract**

Electronic anisotropy was studied for overdoped $(Y,Ca)Ba_2Cu_3O_{7-\delta}$ with various doping levels (p). It was found that the pseudogap-like behavior in the resistivity disappear when p exceeds 0.17, independent of the oxygen deficiency. The anisotropy ratio γ estimated from upper critical fields showed a rapid decrease at around p = 0.18, approaching γ = 3 for p > 0.20.

## 1. Introduction

It is well known that the electronic state of high-$T_c$ superconductors (HTSCs) dramatically changes with carrier concentration. The electronic anisotropy also changes with hole doping from 2 dimensional to 3 dimensional state [1]. However, it is unclear how and where the anisotropy ratio γ reaches the band calculation value in the electronic phase diagram. This links to an important question whether enhanced γ, in other word, two dimensionality is essential for high-$T_c$ superconductivity or not.

To examine this question, $YBa_2Cu_3O_{7-\delta}$ (YBCO) is one of the best systems to be studied because it has the smallest anisotropy in all HTSCs in spite of its relatively high $T_c$ and thus γ can be well determined from the upper critical fields. For the optimal doping, it was reported that $\gamma = (m_c/m_a)^{1/2} = 7.44 \pm 0.25$ [2]. In order to study the overdoped state of YBCO system, we need to substitute Ca for Y. Another purpose of Ca-substitution is to distinguish the effect of oxygen deficiency from the effect of hole concentration change, since it is possible by changing Ca content to control hole concentration and oxygen deficiency independently.

In this study, we measured the in-plane ($\rho_a$) and out-of-plane ($\rho_c$) resistivity and magnetization on $Y_{1-x}Ca_xBa_2Cu_3O_{7-\delta}$ (Ca-YBCO) single crystal with various x and δ values. Anisotropy ratio γ was estimated from the resistivity ratio $\rho_c/\rho_a$ in the normal state and from the upper critical field ratio in the superconducting state. The latter was determined by magnetization and resistive transition measurements. The pseudogap like behaviors [3] in $\rho_a$ and $\rho_c$ disappeared at the doping level p ~ 0.17. It was found that γ rapidly decreased at around p = 0.18 and approaches 3 at p ~ 0.2, which is very close to the band calculation value [4], with keeping a relatively high $T_c$ (70 – 80 K).

## 2. Experiment

$Y_{1-x}Ca_xBa_2Cu_3O_{7-\delta}$ single crystals were grown by a crystal pulling technique, which was described in detail elsewhere [5]. A typical dimension of the as-grown crystals was 5×5×2 mm³. The crystals were detwinned under uniaxial pressure of about 50 MPa and cut into a rectangular shape with a typical size of 1.5×0.7×0.4 mm³ and 0.5×0.5×0.5 mm³ for a-axis and c-axis resistivity measurement, respectively. The samples were annealed in flowing $O_2$ to control oxygen content.

The Ca content was estimated from the relation between optimum $T_c$ value and Ca-content [6]. For determination of oxygen content, the relation between $T_c$ and oxygen deficiency for Ca-YBCO was used [7]. $T_c$ was defined as the onset temperature of diamagnetism for magnetization measurement. We estimated carrier doping level p from the ratio of $T_c/T_{c,max}$, assuming the relationship $T_c/T_{c,max} = 1 - 82.6(p - 0.16)^2$ [8].

The a-axis and c-axis resistivity was measured by a four-probe method, and a-axis resistivity was measured also in magnetic field up to 7 T. A typical current density was about 5 A/cm$^2$. The magnetization was measured with a commercial magnetometer in magnetic field H up to 7 T applied in direction parallel to a-axis and c-axis.

## 3. Results and Discussions

Figure 1 shows the in-plane resistivity $\rho_a$ in zero-field. With increasing doping level, $\rho_a$ systematically decreases. The pseudogap temperature $T^{*(a)}$ is defined as the temperature, at which resistivity starts to deviate from the T-linear behavior. The result of Fig. 1 indicates that $T^{*(a)}$ decreases with increasing p and disappears at p ~ 0.17. This is in agreement with the previous report [3]. Above p ~ 0.18, the temperature dependence becomes weaker at lower T, which can be presented as $\rho_a = \rho_{a0} + AT^\alpha$ ($\alpha$ ~ 1.4 for the sample D3).

The out-of-plane resistivitiy $\rho_c$ also decreases with doping. In addition to the carrier doping effect, we can see the effect of oxygen deficiency on $\rho_c$. For example, though A3 and C1 are at the same doping level, $\rho_c$ of C1 is higher than that of A3 as a result of larger amount of oxygen deficiency. Moreover, C3 shows the lowest resistivity among all the samples in spite of considerable amount of oxygen vacancies ($\delta$ > 0.2). This indicates that the interplane coupling is strong enough in overdoped regime to overcompensate the increase of $\rho_c$ due to oxygen deficiency.

The effect of pseudogap on $\rho_c$ is seen as an increase of resistivity [1]. If we define the pseudogap temperature $T^{*(c)}$ as the temperature at which $\rho_c$ deviates from the T-linear behavior, $T^{*(c)}$ for each sample is nearly identical to $T^{*(a)}$ determined from $\rho_a$ in Fig. 1. Here we see again that $T^{*(c)}$ gradually decreases with doping and disappears at p ~ 0.18.

In order to discuss the anisotropy, we determined upper critical fields $H_{c2}$ from resistivity and magnetization curves. Figure 3 shows the transition behavior of resistivity in magnetic fields H applied parallel to a- and c-axis. For p = 0.170, the resistivity transition shows the typical broadening. Namely, the onset temperature is pinned, while the zero resistivity temperature remarkably goes down with increasing H. For H//c, a sharp drop of $\rho_a$ is observed just above $T_c$, which corresponds to a vortex melting point. These behaviors are typical for HTSC with a wide region of vortex liquid in the vortex phase diagram. On the other hand, it is observed for p = 0.193 that the vortex liquid like behavior is strongly suppressed, and that the resistivity shows parallel shift with increasing magnetic field. This is due to the decrease of anisotropy by overdoping [9,10], which enables us to estimate $T_c(H)$ unambiguously. We evaluated the upper critical fields $H_{c2}$, applying the Werthamer-Helfand-Hohenberg formula [11]. We obtained $H_{c2}^{//a}$ = 130 T, $H_{c2}^{//c}$ = 38 T, $\xi_{ab}$ = 29.4 Å, $\xi_c$ = 8.6 Å, and $\gamma$ = 3.4 for p = 0.193. $\xi_{ab}$ and $\xi_c$ are the coherence lengths in the in-plane and the out-of-plane directions. Since the in-plane anisotropy is small, we assume $\xi_a \sim \xi_b \sim \xi_{ab}$.

We also plotted $T_c(H)$ by magnetization measurement in a similar manner of Ref. [12] and determined $\gamma$, using WHH formula [11]. We also estimated resistivity ratio $\rho_c/\rho_a$ at T = $T_c$ + 3 (K) as a measure of anisotropy. These results are plotted in Fig. 4. Here two of the data points are for the twinned samples, $\gamma$ being evaluated from $\xi_{ab}/\xi_c$. With increasing doping level, $\gamma$ decreases and approaches the value 3 that is the calculation value based on band theory [4]. The relative high value of resistivity ratio results from the oxygen defects effect on $\rho_c$. Therefore, we supposed that the resistivity ratio does not reflect the intrinsic anisotropy for the $CuO_2$ plane.

It is interesting that $\gamma$ rapidly decreases with doping at p ~ 0.18, above which doping dependence is weak. This critical doping level (p ~ 0.18) for the anisotropy change is close to that for the pseudogap disappearance seen in Fig. 1 and 2. This suggests that electronic state drastically changes at around p = 0.18. A rapid change of the electronic state in the overdoped regime is also reported in the measurement of electronic Raman scattering [13]. Although we can conclude that the collapse of pseudogap may trigger an abrupt change in many physical quantities, further studies about detailed mechanism are required.

## 4. Summary

We measured resistivity and magnetization on detwinned Ca-YBCO single crystals with various doping levels to discuss the change of electronic anisotropy. The anisotropy ratio sharply decreases at $p = 0.18$ and approaches the small value of 3 at $p > 0.2$. It may be linked to the disappearance of pseudogap.

## Acknowledgements

This work is supported by New Energy and Industrial Technology Development Organization (NEDO) as Collaborate Research and Development of Fundamental Technologies for Superconductivity Applications.

Table 1 The sample properties measured in this study. Ca concentration (x), critical temperature ($T_c$), maximum critical temperature at optimum doping ($T_{c,max}$), carrier concentration (p) and oxygen vacancy (δ) are shown. Sample A~D were annealed in oxygen atmosphere of 1 atm at 450℃, 500℃, 550℃. Sample E1 was annealed under high-pressure (400atm) oxygen gas. D1 and E1 were not detwinned.

| Sample | x | $T_c$ (K) | $T_{c,max}$ (K) | p | δ |
|---|---|---|---|---|---|
| A1 | 0.00 | 91.5 | 93.5 | 0.144 | 0.160 |
| A2 | 0.00 | 93.5 | 93.5 | 0.160 | 0.120 |
| A3 | 0.00 | 92.7 | 93.5 | 0.170 | 0.095 |
| B1 | 0.03 | 89.2 | 91.0 | 0.175 | 0.120 |
| C1 | 0.12 | 85.7 | 86.5 | 0.170 | 0.245 |
| C2 | 0.12 | 83.0 | 86.5 | 0.182 | 0.215 |
| C3 | 0.12 | 82.0 | 86.5 | 0.185 | 0.207 |
| D1 | 0.14 | 82.0 | 85.5 | 0.182 | 0.240 |
| D2 | 0.14 | 80.0 | 85.5 | 0.188 | 0.225 |
| D3 | 0.14 | 78.0 | 85.5 | 0.193 | 0.212 |
| E1 | 0.12 | 70.0 | 86.5 | 0.208 | 0.150 |

Fig. 1 The temperature dependence of the a-axis resistivity of Ca-YBCO for various doping levels. p = 0.144, 0.160, 0.170, 0.182, 0.185, 0.188 and 0.193. The arrows indicate pseudogap temperature $T^{*(a)}$. The error of resistivity value is about 5%.

Fig. 2 The temperature dependence of the c-axis resistivity of Ca-YBCO for various doping levels and oxygen contents. (p, δ) = (0.144, 0.128), (0.160, 0.096), (0.170, 0.196), (0.170, 0.076), (0.182, 0.172) and (0.185, 0.166). The arrows indicate pseudogap temperature $T^{*(c)}$.

Fig. 3 Resistive transition in various magnetic fields from 0 to 7 T. (a)for Sample A3 (p = 0.170) in H//a, (b)for p = 0.170 in H//c, (c)for Sample D3 (p = 0.193) in H//a and (d)for p = 0.193 in H//c.

Fig. 4 Anisotropy ratio γ as a function of the doping level p. The crosses show resistivity ratio $(\rho_c/\rho_a)^{1/2}$ at $T = T_c + 3$ (K). The triangles show upper critical field ratio $H_{c2}^{//a}/H_{c2}^{//c}$ estimated from resistive transition in magnetic fields, using WHH formula [11]. The squares represent $H_{c2}^{//a}/H_{c2}^{//c}$ estimated from magnetization.

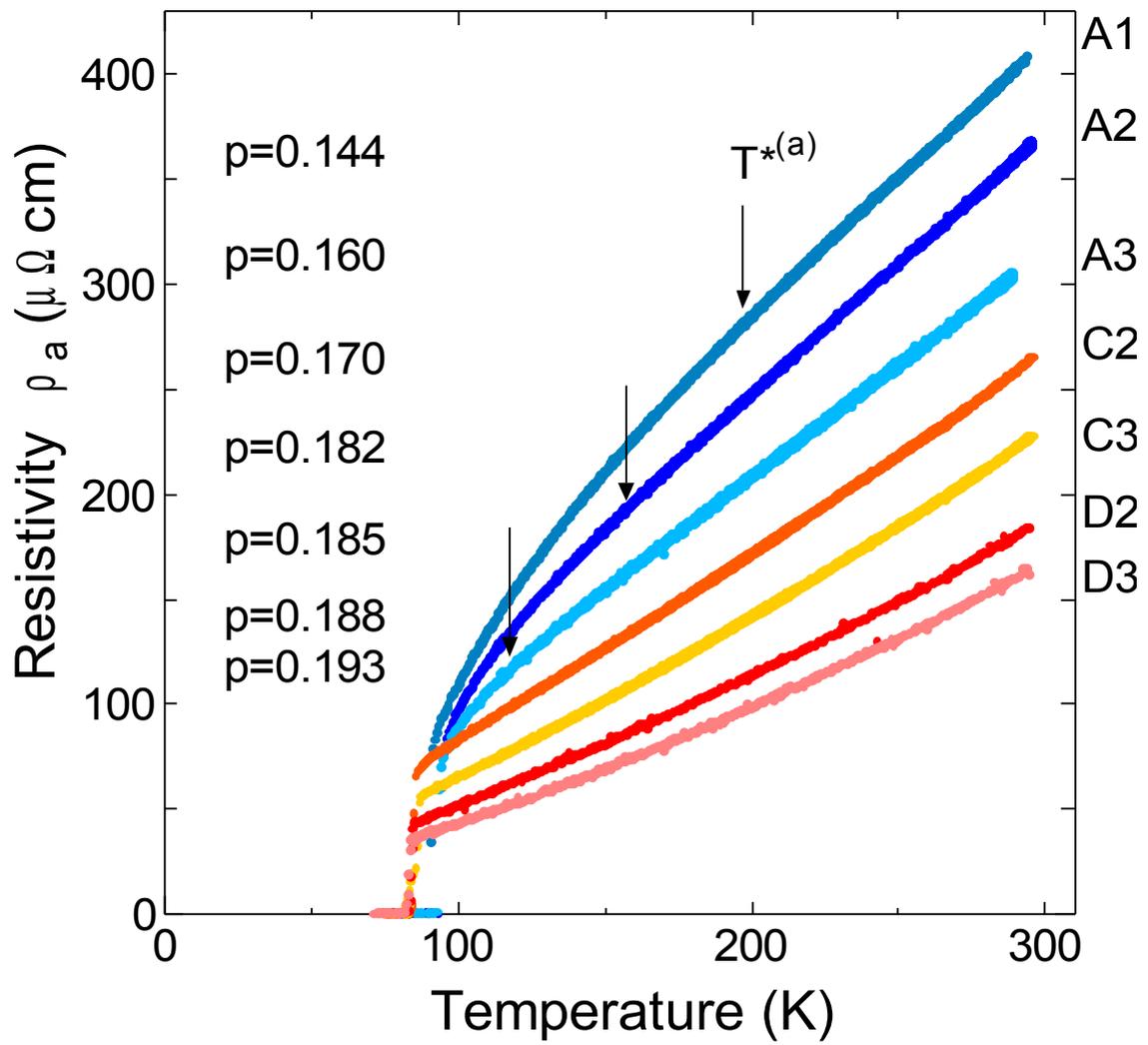

Fig. 1

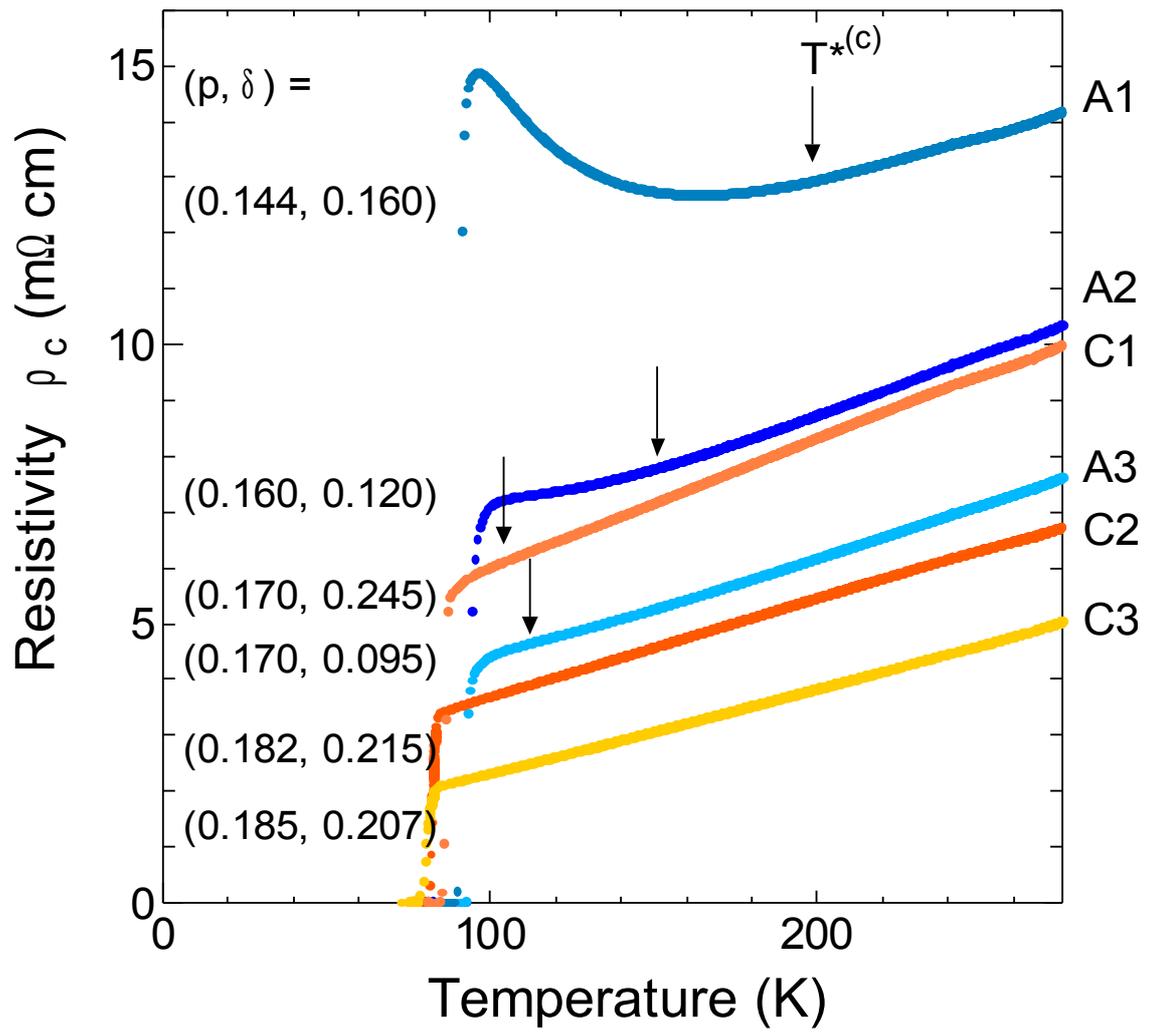

Fig. 2

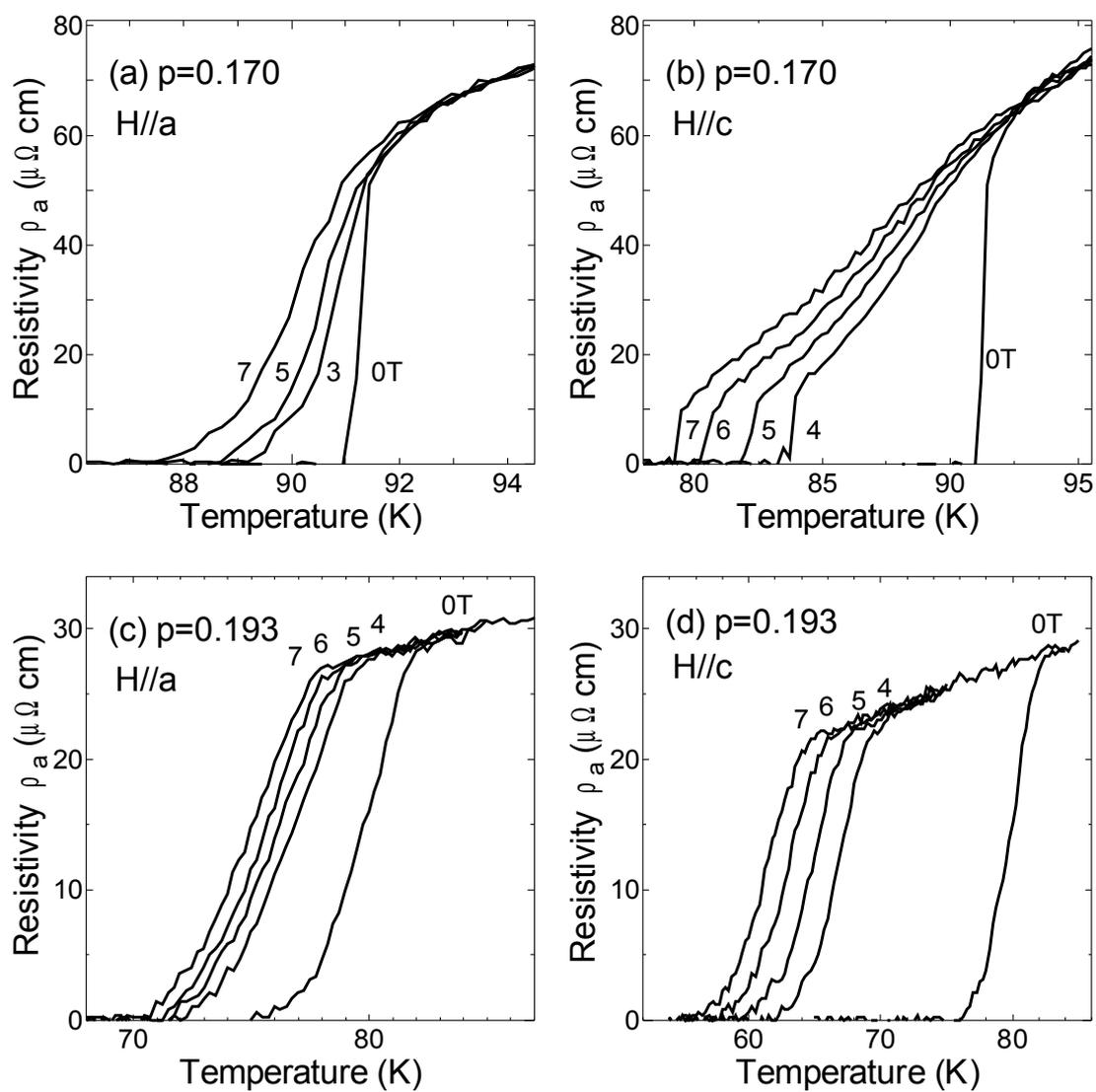

Fig. 3

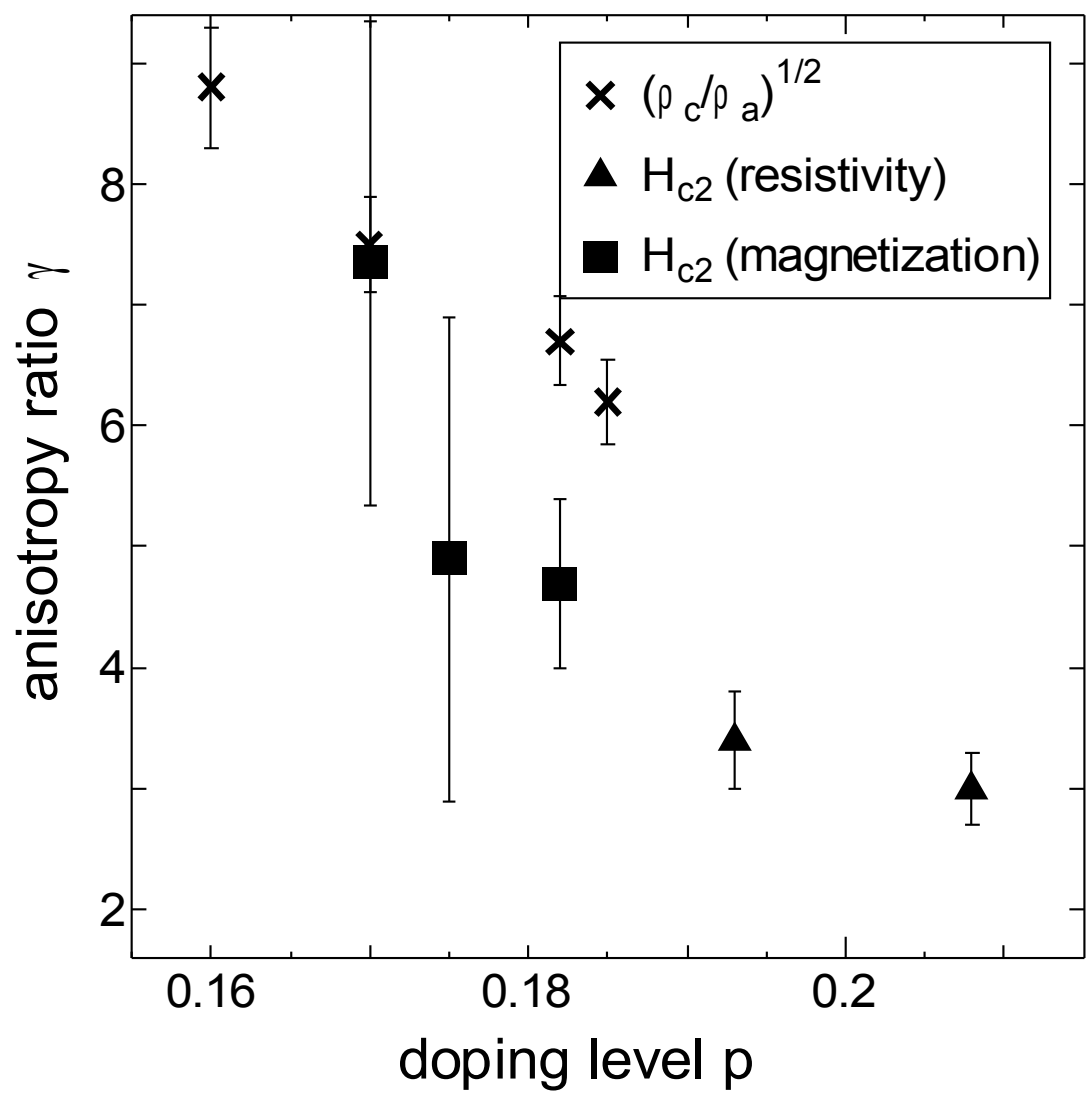

Fig. 4